\documentclass{Interspeech2024}
\usepackage{multirow}

\interspeechcameraready

\title{Zero-Shot Fake Video Detection by Audio-Visual Consistency}

\name[affiliation={1}]{Xiaolou}{Li}
\name[affiliation={1}]{Zehua}{Liu}
\name[affiliation={2}]{Chen}{Chen}
\name[affiliation={1}]{Lantian}{Li}
\name[affiliation={1}]{Li}{Guo}
\name[affiliation={2}]{Dong}{Wang}

\address{
  $^1$School of Artificial Intelligence, Beijing University of Posts and Telecommunications, China \\
  $^2$Center for Speech and Language Technologies, BNRist, Tsinghua University, China
  \thanks{This work was supported by the National Natural Science Foundation of China (NSFC) under Grants No.62301075/62171250. X. Li and Z. Liu are joint first authors.}
  }
\email{Corresponding authors:~lilt@bupt.edu.cn, wangdong99@mails.tsinghua.edu.cn}


\begin{document}

\maketitle

\begin{abstract}

Recent studies have advocated the detection of fake videos as a one-class detection task, 
predicated on the hypothesis that the consistency between audio and visual modalities of genuine data is more significant than that of fake data. 
This methodology, which solely relies on genuine audio-visual data while negating the need for forged counterparts, 
is thus delineated as a `zero-shot' detection paradigm. 
This paper introduces a novel zero-shot detection approach anchored in content consistency across audio and video. 
By employing pre-trained ASR and VSR models, we recognize the audio and video content sequences, respectively.
Then, the edit distance between the two sequences is computed to assess whether the claimed video is genuine. 
Experimental results indicate that, compared to two mainstream approaches based on semantic consistency and temporal consistency, 
our approach achieves superior generalizability across various deepfake techniques and demonstrates strong robustness against audio-visual perturbations. 
Finally, state-of-the-art performance gains can be achieved by simply integrating the decision scores of these three systems.

\end{abstract}

\section{Introduction}


In recent years, the development of deepfake technologies has made it possible to generate high-fidelity fake videos ~\cite{westerlund2019emergence,mahmud2021deep,kwok2021deepfake,sharma2022review}. These technologies leverage advanced methods such as face-swapping, lip-syncing for video generation, and speech synthesis or voice conversion for audio generation.
These fake videos pose significant risks of misleading the public, damaging reputations, threatening security, and undermining trust~\cite{de_Rancourt-Raymond_Smaili_2023,singh2023exploding,choudhary2023unmasking}. Consequently, developing deepfake detection technologies has emerged as a critical concern.

As deepfake advances, the development of countermeasures has concurrently evolved~\cite{rana2022deepfake,yu2021survey,almutairi2022review,seow2022comprehensive,deshmukh2020deepfake}. Initially, deepfakes primarily relied on face-swapping techniques, producing fake videos characterized by unnaturally smooth areas within frames or notable discontinuities between frames. Consequently, early fake video detection strategies identified these forgeries by recognizing such artifacts. For example, Zheng et al.~\cite{zheng2021exploring} proposed detecting forgeries by capturing the discontinuities between video frames. Haliassos et al.~\cite{haliassos2021lips} achieved deepfake detection by identifying semantic irregularities in the lip movements within videos.

However, with the further advancement of deepfake technologies, relying solely on the video modality for fake video detection has become exceedingly challenging. To address this challenge, researchers expanded their focus to introduce audio modality to assist in fake video detection, leading to audio-visual multi-modal forgery detection. Initially, researchers adopted an end-to-end binary classification framework to discriminate between genuine and fake videos. For instance, Wang et al.~\cite{wang2022audio} proposed a multi-modal detection network that takes raw audio and video streams as input. By leveraging an attention mechanism, this network integrates audio and video features deeply, ultimately distinguishing between genuine and fake videos using a binary classifier.

Although these approaches demonstrated preliminary effectiveness, their fundamental limitation was the independent detection of artifacts in audio and video without considering the impact of deepfakes on the consistency between audio and video. Genuine videos naturally possess intrinsic consistency between audio and video modalities, while deepfakes may somewhat corrupt this consistency. Therefore, several studies have focused on evaluating the consistency between audio and video for deepfake detection. For example, Shahzad et al.~\cite{shahzad2022lip} detected forgeries by quantifying the mismatch between the lip sequence extracted from the video and the synthetic lip sequence generated from the audio by the Wav2Lip model~
\cite{prajwal2020lip}. Chugh et al.~\cite{chugh2020not} introduced a contrastive loss to enforce similarity in audio and video representations of genuine video pairs and dissimilarity in those of fake pairs, thereby establishing inter-modality similarity. Zhang et al.~\cite{zhang2024joint} adopted the same strategy. Cheng et al.~\cite{cheng2023voice} argued that there is a high homogeneity between a person's face and voice. They, therefore, detect fake videos by assessing the matching degree between face and voice representations.

\begin{figure*}[!htb]
  \centering
  \vspace{-2mm}
  \includegraphics[width=0.92\linewidth]{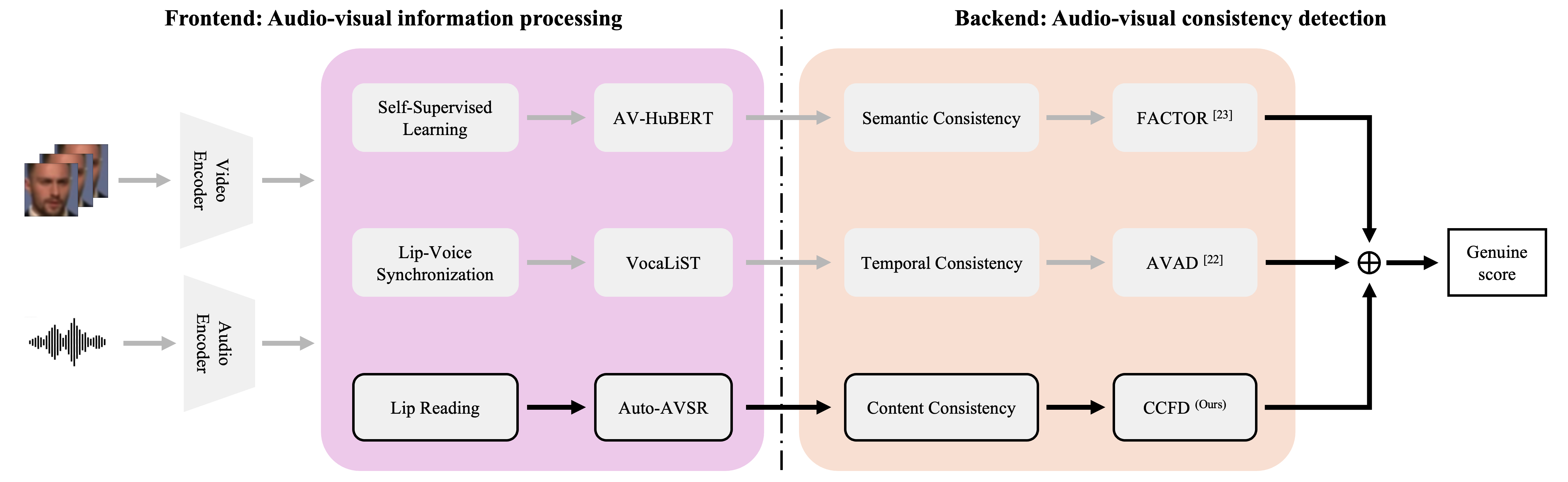}
  \vspace{-1mm}
  \caption{The unified framework of zero-shot fake video detection.}
  \label{fig:framework}
\vspace{-3mm}
\end{figure*}

Despite the effectiveness of incorporating audio-visual consistency in improving detection performance, these methods generally rely on an end-to-end two-class classification framework. This framework performs well for detecting \emph{specific} deepfakes but lacks generalizability for unseen deepfakes. Considering that the consistency of audio-visual modalities is an intrinsic property of genuine videos, this two-class classification task can be re-conceptualized as a one-class detection task that solely detects whether a video is genuine. Notably, this one-class framework requires only genuine audio-visual data for modelling, without the need for any fake data, thus regarded as `zero-shot' deepfake detection. For instance, Cozzolino et al.~\cite{cozzolino2023audio} utilized face recognition and speaker recognition models trained on genuine data to detect forgeries by assessing the consistency between face identity and speaker identity. Feng et al.~\cite{feng2023self} trained solely on genuine audio-visual data and detected forgeries by assessing the temporal synchrony between audio and video. Tal et al.~\cite{reiss2023detecting} used the AV-HuBERT model~\cite{shi2022learning} trained on real audio-visual data to detect forgeries by quantifying the distance between semantic representations of audio and video. Since these zero-shot approaches only require genuine data for modelling, they can detect \emph{any} type of deepfakes, demonstrating stronger generalization capabilities.

This paper introduces a novel audio-visual deepfake detection method based on voice-lip content consistency. The fundamental assumption of our method is that genuine data possess intrinsic content consistency between voice streams and lip movements. To this end, we first employ automatic speech recognition (ASR) and visual speech recognition (VSR) models trained on genuine data to decode the content sequences from audio and video, respectively. We then compute the edit distance between the two content sequences as a metric to evaluate the degree of content consistency between the modalities. Experimental results demonstrate that compared to semantic-consistency-based FACTOR~\cite{reiss2023detecting} and temporal-consistency-based AVAD~\cite{feng2023self}, our content-consistency method achieves great performance across a variety of deepfake datasets (including FakeAVCeleb~\cite{khalid2021fakeavceleb} and DeepFakeTIMIT~\cite{korshunov2018deepfakes}), showcasing superior generalizability. Moreover, we ascertain that our method is more robust by introducing various perturbations into both audio and video. Finally, believing in the complementary strengths of the three distinct consistency approaches,
we advocate for a simple score fusion approach to combine these methods. Our results indicate that this fusion achieves state-of-the-art (SOTA) performance in deepfake detection, setting a new benchmark in the field.


\section{Zero-Shot Fake Video Detection}
\label{sec:framework}

In this section, we delineate a unified framework for zero-shot fake video detection, as shown in Figure~\ref{fig:framework}. 
This framework is composed of two components: frontend audio-visual information processing and 
backend audio-visual consistency detection.

\subsection{Frontend: Audio-visual information processing}

At the frontend, various audio-visual information processing tasks employ different model architectures, training paradigms, and objectives. 
Despite these differences, the underlying methodology remains consistent, 
involving encoding genuine audio and visual inputs to derive their respective latent representations, then establishing the correlation between the representations of the two modalities, and finally leveraging the correlation to learn the pretext task. 
For instance, AV-HuBERT~\cite{shi2022learning} adopts a self-supervised learning strategy, engaging in an iterative process of feature clustering and 
learning new features via a masked prediction loss. 
This process uncovers strong correlations between audio streams and lip movements, 
yielding a highly effective pre-trained model that has been successfully deployed in various audio-visual downstream applications. 
Besides, VocaLiST~\cite{kadandale2022vocalist} designs a powerful cross-modal Transformer model for learning the correlation across audio and visual streams. 
Then, it outputs a score indicating whether the voice and lip motion are synchronised.
Moreover, AV-ASR~\cite{ma2021end, ma2023auto}, by leveraging the correlation of contextual information across audio and visual streams, 
integrates audio and visual contextual representations to enhance visual speech recognition, also known as lip reading.

\subsection{Backend: Audio-visual consistency detection}

Considering that these audio-visual information processing frontends require only genuine audio-visual data during the training phase without the need for any fake data, 
these well-trained frontend models have proficiently learned the intrinsic inter-modality correlations within genuine data.
A natural intuition is that these correlations observed between audio and video modalities in genuine data are much weaker in fake data.
Hence, at the backend, quantifying these correlations allows for the measurement of consistency across audio and video elements, 
facilitating the determination of a video's authenticity. 
For instance, FACTOR~\cite{reiss2023detecting} leverages a pre-trained AV-HuBERT model to extract latent representations of audio and video; 
it then employs cosine similarity to assess the semantic consistency between these representations, yielding a decision score. 
This approach represents a semantic-consistency-based method of fake video detection and has achieved great performance on the FakeAVCeleb dataset. 
In addition, AVAD~\cite{feng2023self} trained a model on lip-voice synchronization, generating features that describe the temporal synchronization between audio and visual streams, 
and subsequently predicts a consistency score based on these features, thus representing a temporal-consistency-based approach to fake video detection.

In this paper, we introduce a novel fake video detection approach based on content consistency, 
termed \textbf{c}ontent \textbf{c}onsistency \textbf{f}ake \textbf{d}etection, CCFD. 
Specifically, we posit that for genuine data, there is a strong correlation between the content information of audio and visual streams. 
Following this hypothesis, we leverage ASR and VSR models within the AV-ASR framework to decode the content sequences of audio streams and lip movements, respectively. 
The consistency between audio and video is then measured by computing the edit distance between these two content sequences. 
In this study, we use the audio content sequence decoded by ASR as the reference and the lip content sequence decoded by VSR as the hypothesis when computing the word error rate (WER), 
thereby determining if the claimed video is genuine or not.


Finally, we believe various consistency-based detection methods, grounded in different tasks and assumptions, possess inherent complementary qualities. 
Therefore, fusing the decision scores output by different detection methods is feasible to arrive at an improved detection assessment.

\section{Experiment Settings}
\label{sec:setting}

\subsection{Data}
\vspace{-0.1cm}

Our experiments used two datasets: FakeAVCeleb~\cite{khalid2021fakeavceleb} and DeepFakeTIMIT~\cite{korshunov2018deepfakes}.

\emph{FakeAVCeleb}, a large-scale audio-visual deepfake dataset. The genuine videos were selected from VoxCeleb2~\cite{chung2018voxceleb2}.
It employed face-swapping algorithms such as Faceswap~\cite{korshunova2017fast} and FSGAN~\cite{nirkin2019fsgan} to generate swapped fake videos.
Besides, it used an SV2TTS tool~\cite{jia2018transfer} to generate cloned audios.
After generating fake videos and audios, Wav2Lip~\cite{prajwal2020lip} was applied to fake videos to reenact the videos based on fake audios. 
In our experiments, we sampled 50 genuine videos and 2,085 fake videos from 50 celebrities for performance evaluation.

\emph{DeepFakeTIMIT}, a standard deepfake dataset. 
It encompasses 320 genuine videos selected from VidTIMIT\footnote{http://conradsanderson.id.au/vidtimit/}, 
featuring 16 pairs of speakers with similar visual characteristics.
Furthermore, 320 fake videos were produced using advanced face-swapping techniques. 

In our experiments, all videos were transcoded to a frame rate of 25 frames per second, 
and all audios were resampled to a sampling rate of 16kHz.

\subsection{Systems}

\subsubsection{SCFD: Semantic-consistency fake detection}

Followed by FACTOR\footnote{https://github.com/talreiss/FACTOR}, we constructed a semantic-consistency fake detection (SCFD) system.

Firstly, we followed the preprocessing procedure outlined by Auto-AVSR\footnote{https://github.com/mpc001/auto\_avsr/tree/main/preparation}, which includes: (1) Utilizing RetinaFace~\cite{deng2020retinaface} for facial landmark detection. (2) Affine transformation is applied to align and stabilize the facial region in the original video, reducing the impact of head movements and centring the mouth area. (3) After alignment and stabilization, a 96x96 pixel region centred around the mouth is cropped from the frames.

Subsequently, semantic representations for each video frame and its corresponding audio segment are independently extracted using the video and audio encoders provided by AV-HuBERT. 
The cosine similarity between the two semantic representations is computed, resulting in a semantic consistency score for each frame.

Finally, we take the 3rd percentile of scores from all frames as the result to obtain a video-level semantic consistency score.

\subsubsection{TCFD: Temporal-consistency fake detection}

Inspired by AVAD~\cite{feng2023self}, we developed a temporal-consistency fake detection (TCFD) system.

Initially, we adhere to the preprocessing protocol established by VocaLiST\footnote{https://github.com/vskadandale/vocalist}, 
utilizing facial detection techniques to extract the facial region from the original videos. After resizing the extracted region to 96x96 pixels, we specifically crop the area encompassing the lips.



Subsequently, employing a window length of five frames with a stride of one frame, the lip stream and audio stream are concurrently input into the VocaLiST pre-trained model. This procedure calculates a synchronization score for each window, reflecting the temporal alignment between the audio and video streams within that specific five-frame window. 

Ultimately, to determine the video's overall temporal consistency, the average synchronization score across all windows is computed, providing a comprehensive video-level temporal consistency score.

\vspace{-0.1cm}
\subsubsection{CCFD: Content-consistency fake detection}

In this study, we have introduced a content-consistency fake detection (CCFD) system to detect fake videos 
by assessing the content consistency between audio and visual streams.

The pipeline of data preprocessing is consistent with that of SCFD. 
We utilize Visual Speech Recognition (VSR) and Automatic Speech Recognition (ASR) models ~\cite{gulati2020conformer, he2016deep, stafylakis2017combining} that 
have been released in the Auto-AVSR repository\footnote{https://github.com/mpc001/auto\_avsr/tree/main}. 
The decoding process~\cite{petridis2018audio} is conducted using the BeamSearch algorithm, setting the beam size to 40.

Subsequently, video and audio streams are separately fed into the VSR and ASR models to decode the respective content sequences.
Considering the superior accuracy of ASR over VSR in recognizing content, 
we designate the audio content sequence decoded by ASR as the reference and the lip content sequence decoded by VSR as the hypothesis. 
The degree of content consistency between these sequences is quantified by computing the Word Error Rate (WER), 
thereby offering a metric to measure the authenticity of the video.

\vspace{-0.1cm}
\subsubsection{System fusion}

Intuitively, the three fake detection systems leverage different consistency criteria, suggesting inherent complementarity among them.
Therefore, we believe that fusing the output of these systems could significantly enhance the generalizability and robustness of the final detection. 

We simply average the scores from the three systems to achieve this fusion.
Before this fusion, it is essential to normalize the score from each system to standardize the value range. 
In our experiments, the min-max normalization method is utilized for SCFD and TCFD systems. 
For the CCFD system, score normalization is implemented using the formula 1-min(WER, 1). 
This normalization process ensures that the scores across different systems are comparable, facilitating a balanced and effective fusion of their outputs.

\begin{table*}[!htb]
  \vspace{-2mm}
  \caption{AUC results for fake detection systems across datasets.}
  \vspace{-2mm}
  \label{tab:general}
  \centering
  \resizebox{1.7\columnwidth}{!}{
  \begin{tabular}{ccccccccccc}
    \toprule
        \multirow{2}{*}{Dataset}          & \multicolumn{7}{c}{FakeAVCeleb}         & \multirow{3}{*}{DeepFakeTIMIT} & \multirow{3}{*}{Mean} & \multirow{3}{*}{Std.}    \\
        \cmidrule(r){2-8} 
                                          & \multirow{2}{*}{RVFA}   &   \multicolumn{3}{c}{FVRA}  & \multicolumn{3}{c}{FVFA}    &       &        &          \\
        \cmidrule(r){3-5} \cmidrule(r){6-8}
        Method                            &           & WL       & GAN     & FS      & WL     & GAN-WL    & FS-WL       &         &         &   \\ 
        \cmidrule(r){1-1} \cmidrule(r){2-8} \cmidrule(r){9-9} \cmidrule(r){10-11}
        SCFD        & \bf{0.9924} & 0.9481  & 0.7741  & 0.7167  & 0.9686  & 0.9687  & 0.9656  & 0.9110  & 0.9056  & 0.0961    \\ 
        TCFD        & 0.7824  & 0.9545   & 0.5262   & 0.4887   & 0.9728  & 0.9734  & 0.9685  & 0.8144  & 0.8101  & 0.1883      \\ 
        CCFD        & 0.6880  & 0.9575  & \bf{0.8403}  & \bf{0.8424}  & 0.9522   & 0.9472  & 0.9412  & 0.9315  & 0.8875  & 0.0877  \\ 
        Fusion      & 0.8624  & \bf{0.9811}  & 0.8144  & 0.7704  & \bf{0.9857}  & \bf{0.9841}  & \bf{0.9799}  & \bf{0.9786}  & \bf{0.9196}  & \bf{0.0837}   \\ 
    \bottomrule
  \end{tabular}}
  \vspace{-2mm}
\end{table*}

\section{Experimental Results}
\label{sec:result}

In this section, we evaluate the generalizability and robustness of various fake detection systems, 
emphasizing the different impact of audio-visual consistency criteria.

\subsection{Generalization tests}

For generalization tests, experiments were conducted on the FakeAVCeleb and DeepFakeTIMIT datasets. 
Given the variety of deepfake techniques in FakeAVCeleb, we split this dataset into several subsets based on the deepfake mode and technique for detailed analysis and reported performance on each subset. 
Deepfake modes were categorized into three groups: RVFA (real video with fake audio), FVRA (fake video with real audio), and FVFA (fake video and audio), 
and deepfake techniques include Wav2Lip (WL), FSGAN (GAN), and FaceSwap (FS).
The Area Under the Curve (AUC) was employed as the evaluation metric. 
We measured the mean and standard deviation of the AUC scores on different datasets and used these quantities to evaluate the generalizability of a fake detection method.
Notably, higher means and lower standard deviations indicate superior generalizability. The results are reported in Table~\ref{tab:general}.

Firstly, SCFD achieved the highest mean AUC among the three consistency-based detection systems, while CCFD exhibited the smallest standard deviation. 
More careful analysis revealed that SCFD was most effective in the RVFA mode; TCFD excelled against WL-based techniques; 
and CCFD demonstrated exceptional efficacy against GAN- and FS-based techniques within the FVRA mode, 
highlighting a clear bias of consistency criteria towards specific deepfake modes and techniques.

Secondly, our proposed system, CCFD, showed remarkable stability across datasets except for RVFA. 
This can be attributed to the high fidelity of the audio synthesized by SV2TTS, which permitted the ASR model to achieve accurate results even on fake audio.

Finally, fusing the three systems resulted in enhanced accuracy and generalizability, 
underscoring the complementary nature of the consistency criteria in the realm of fake detection.

\vspace{-2mm}
\subsection{Robustness tests}
\vspace{-1mm}

In practical applications, videos often encounter various noises and corruptions, such as background noise and compression artifacts, which may affect the performance of fake detection systems. Thus, evaluating the robustness of a fake detection system against a range of perturbations is of paramount importance.

\begin{figure}[!htp]
  \centering
  \includegraphics[width=\linewidth]{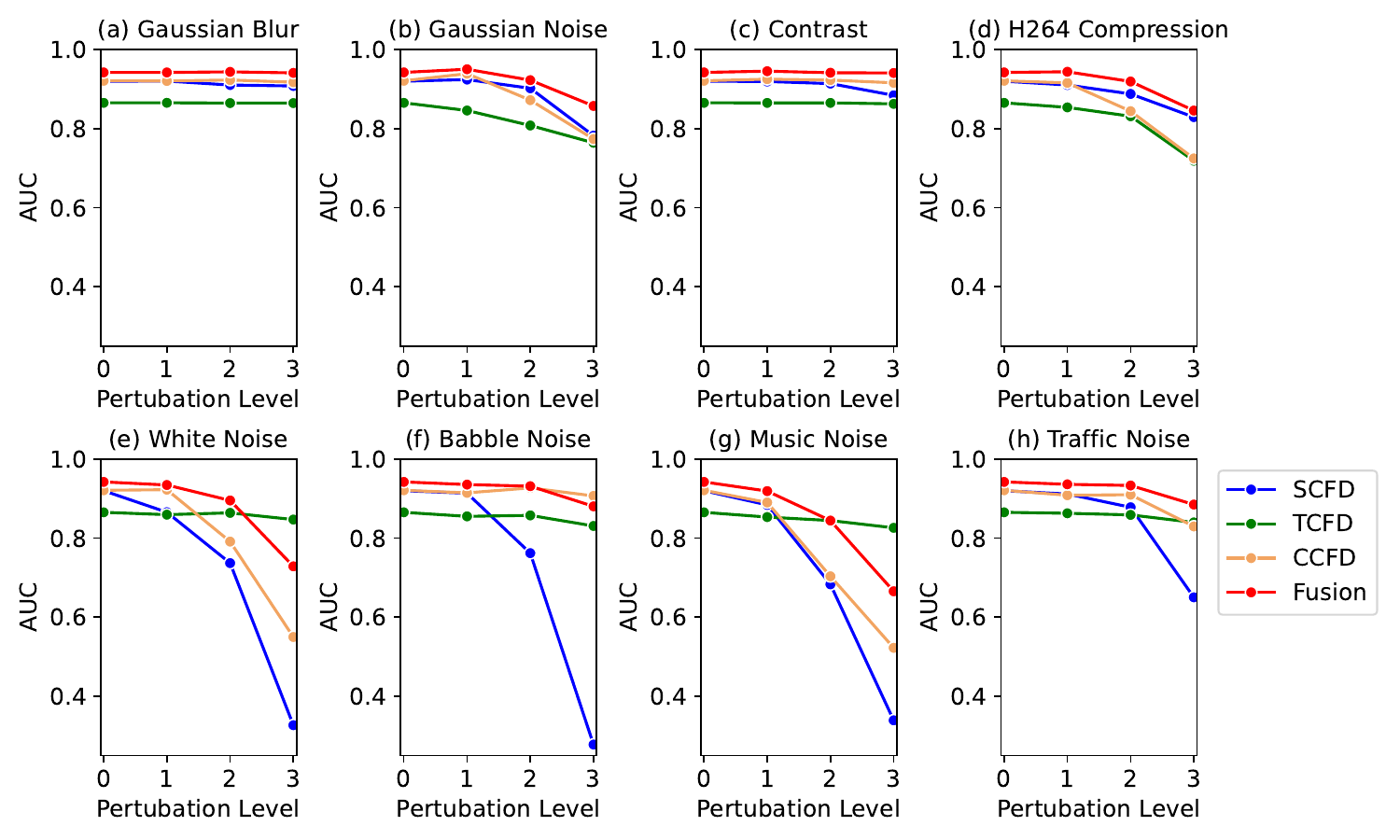}
  \vspace{-4mm}
  \caption{AUC on FakeAVCeleb under levels of perturbations.}
  \label{fig:robust}
  \vspace{-4mm}
\end{figure}

We implemented three levels of video perturbations using the Kornia library\footnote{https://github.com/kornia/kornia} and FFmpeg library\footnote{https://ffmpeg.org/}, with the specifics detailed in Table~\ref{tab:perturb}. For audio perturbations, four types of noise were added using the Torchaudio library\footnote{https://pytorch.org/audio/stable/index.html} at three signal-to-noise ratio (SNR) levels: 12.5 dB, 2.5 dB, and -7.5 dB. The results on FakeAVCeleb are depicted in Figure~\ref{fig:robust} and summarized in Table~\ref{tab:robust}.

\begin{table}[!htb]
  \vspace{-1mm}
  \caption{Three levels of video perturbations.}
  \vspace{-2mm}
  \label{tab:perturb}
  \centering
  \resizebox{0.8\columnwidth}{!}{
  \begin{tabular}{lcccc}
    \toprule
        Type      & Blur  & Noise & Contrast & Compression \\ 
        Parameter & sigma & std   & factor   & CRF         \\ 
    \midrule
        Level 1   & 0.1   & 0.01  & 0.8      & 33          \\ 
        Level 2   & 2     & 0.05  & 1.2      & 40          \\ 
        Level 3   & 5     & 0.1   & 2        & 47          \\ 
    \bottomrule
  \end{tabular}}
  \vspace{-2mm}
\end{table}

\begin{table}[!htp]
  \centering
  \caption{Mean and Std. of different systems in robustness tests.}
  \vspace{-2mm}
  \label{tab:robust}
  \resizebox{0.6\columnwidth}{!}{
  \begin{tabular}{ccccc}
    \toprule
        \multirow{2}{*}{AUC}   & \multicolumn{2}{c}{Video} &  \multicolumn{2}{c}{Audio} \\
        \cmidrule(r){2-5} 
                               & Mean         & Std.       & Mean       & Std.          \\
        \cmidrule(r){1-1} \cmidrule(r){2-5} 
        SCFD                    & 0.894       & 0.039      & 0.704      & 0.230      \\ 
        TCFD                    & 0.838       & 0.043      & 0.851      & \bf{0.012} \\ 
        CCFD                    & 0.888       & 0.062      & 0.823      & 0.137      \\ 
        Fusion                  & \bf{0.927}  & \bf{0.032} & \bf{0.879} & 0.083      \\ 
    \bottomrule
  \end{tabular}}
\end{table}

It can be observed that SCFD demonstrated considerable robustness against video perturbations but was highly sensitive to audio perturbations. This implies a significant alteration in the audio representations that the AV-HuBERT model extracts upon adding audio perturbations. In contrast, TCFD showed an inverse trend, demonstrating robustness to audio perturbations and vulnerability to video perturbations. This indicates that the audio-visual synchronization detection model lacks robustness against video perturbations. The performance of our proposed CCFD lies between SCFD and TCFD, showing stability in the face of both audio and video perturbations. 
After integrating the three systems, a consistent advantage in robustness was achieved across all test cases. This outcome reaffirms the complementary nature of the three consistency criteria, which can be combined to construct a stronger fake detection system.

\section{Conclusion}
\label{sec:conc}

This paper focuses on zero-shot fake video detection. 
We introduce a unified framework for zero-shot fake detection methods: an audio-visual information processing frontend and an audio-visual consistency detection backend. 
Following this, we constructed three zero-shot fake detection systems with different consistency criteria, 
including a novel method based on content consistency. 
Experimental results demonstrate that different systems excel at different types of deepfakes and are sensitive to different audio/video perturbations.
Compared to existing methods based on temporal consistency and semantic consistency, our proposed content consistency detection system presents stable generalizability and robustness. 
By fusing these systems, we achieved SOTA performance on the FakeAVCeleb dataset, highlighting the complementarity among the three consistency criteria. 
Future work will continue exploring more consistency criteria for our zero-shot fake detection framework.


\bibliographystyle{IEEEtran}
\bibliography{mybib}

\end{document}